\documentclass[twocolumn,english]{revtex4-2}
\usepackage[T1]{fontenc}
\usepackage[latin9]{inputenc}
\usepackage{color}
\usepackage{babel}
\usepackage{units}
\usepackage{amsmath}
\usepackage{amssymb}
\usepackage{graphicx}
\usepackage{rotfloat}
\usepackage[unicode=true,pdfusetitle,
 bookmarks=true,bookmarksnumbered=false,bookmarksopen=false,
 breaklinks=false,pdfborder={0 0 1},backref=false,colorlinks=true]
 {hyperref}
\hypersetup{
 citecolor=blue}

\makeatletter
\usepackage{chngcntr}
\counterwithout{figure}{section}
\counterwithout{equation}{section}
\usepackage{times}
\usepackage{graphicx}
\usepackage{float}
\usepackage{stmaryrd}

\makeatother

\begin{document}

\global\long\def\theenumi{\alph{enumi}}%

\global\long\def\rmi{\mathbf{\textrm{i}}}%

\global\long\def\rme{\mathbf{\textrm{e}}}%

\global\long\def\rmd{\mathbf{\textrm{d}}}%

\global\long\def\sgn{\mathrm{sign}}%

\global\long\def\id{\mathbf{1}}%

\global\long\def\C{\mathbb{C}}%

\global\long\def\R{\mathbb{R}}%

\global\long\def\Q{\mathbb{Q}}%

\global\long\def\N{\mathbb{N}}%

\global\long\def\Z{\mathbb{Z}}%

\global\long\def\Id{\mathbf{\mathbb{I}}}%

\global\long\def\V{\mathcal{V}}%

\global\long\def\set#1#2{\left\{  #1\thinspace:\thinspace#2\right\}  }%


\global\long\def\Nz{\N_{0}}%

\global\long\def\Aa{\mathcal{A}}%

\global\long\def\tree{\mathcal{T}_{\alpha}}%

\global\long\def\graph{\mathcal{G}_{\alpha}}%

\global\long\def\pl{\gamma_{L}}%

\global\long\def\pr{\gamma_{R}}%

\global\long\def\paths{\Gamma_{\alpha}}%

\global\long\def\emap{E_{\alpha}}%

\global\long\def\tA{\mathit{A}}%

\global\long\def\tB{\mathit{B}}%

\global\long\def\tC{\mathit{G}}%

\global\long\def\modd{\mathop{mod^{*}} }%

\global\long\def\mdd#1#2{\left[#1\right]_{#2}^{*} }%

\global\long\def\nA{z_{A}\emph{}}%

\global\long\def\nB{z_{B}\emph{}}%

\global\long\def\NA#1{Z_{A}(#1)\emph{}}%

\global\long\def\NB#1{Z_{A}(#1)\emph{}}%

\global\long\def\QQ#1{Q_{#1}\emph{}}%

\global\long\def\QQQ#1{\widetilde{Q}_{#1}\emph{}}%

\global\long\def\ind#1{\mathfrak{i}_{#1}\emph{}}%

\global\long\def\col#1{\mathfrak{c}_{#1}\emph{}}%

\global\long\def\IDS{N_{\alpha}}%

\global\long\def\spec#1{\emph{\ensuremath{\mathrm{Spec}\left(#1\right)}}}%

\global\long\def\co{\textbf{c}}%

\global\long\def\cop{\textbf{\ensuremath{\widetilde{\co}}}}%

\global\long\def\cm{[\textbf{c},m]}%

\global\long\def\cmn{[\textbf{c},m,n]}%

\global\long\def\alk{\alpha_{k}}%

\global\long\def\Co{\mathscr{C}}%

\global\long\def\Ham{H_{\alpha}}%

\global\long\def\Hrat{H_{\frac{p}{q}}}%

\global\long\def\Halk{H_{\alk}}%

\global\long\def\sigc{\sigma_{\co}}%

\global\long\def\sigcm{\sigma_{[\co,m]}}%

\global\long\def\sigcmn{\sigma_{[\co,m,n]}}%

\title{From Quasiperiodicity to a Complete Coloring of the Kohmoto Butterfly}
\author{Ram Band$^{1}$ and Siegfried Beckus$^{2}$}
\affiliation{$^{1}$Department of Mathematics, Technion--Israel Institute of Technology,
Haifa, Israel~\\
$^{2}$Institute of Mathematics, University of Potsdam, Potsdam, Germany}
\begin{abstract}
The spectra of the Kohmoto model give rise to a fractal phase diagram,
known as the Kohmoto butterfly. The butterfly encapsulates the spectra
of all periodic Kohmoto Hamiltonians, whose index invariants are sought
after. Topological methods -- such as Chern numbers -- are ill defined
due to the discontinuous potential, and hence fail to provide index
invariants. This Letter overcomes that obstacle and provides a complete
classification of the Kohmoto model indices. Our approach encodes
the Kohmoto butterfly as a spectral tree graph, reflecting the quasiperiodic
nature via the periodic spectra. This yields a complete coloring of
the phase diagram and a new perspective on other spectral butterflies.
\end{abstract}
\maketitle
\begin{figure}
\centering
\includegraphics[width=17cm]{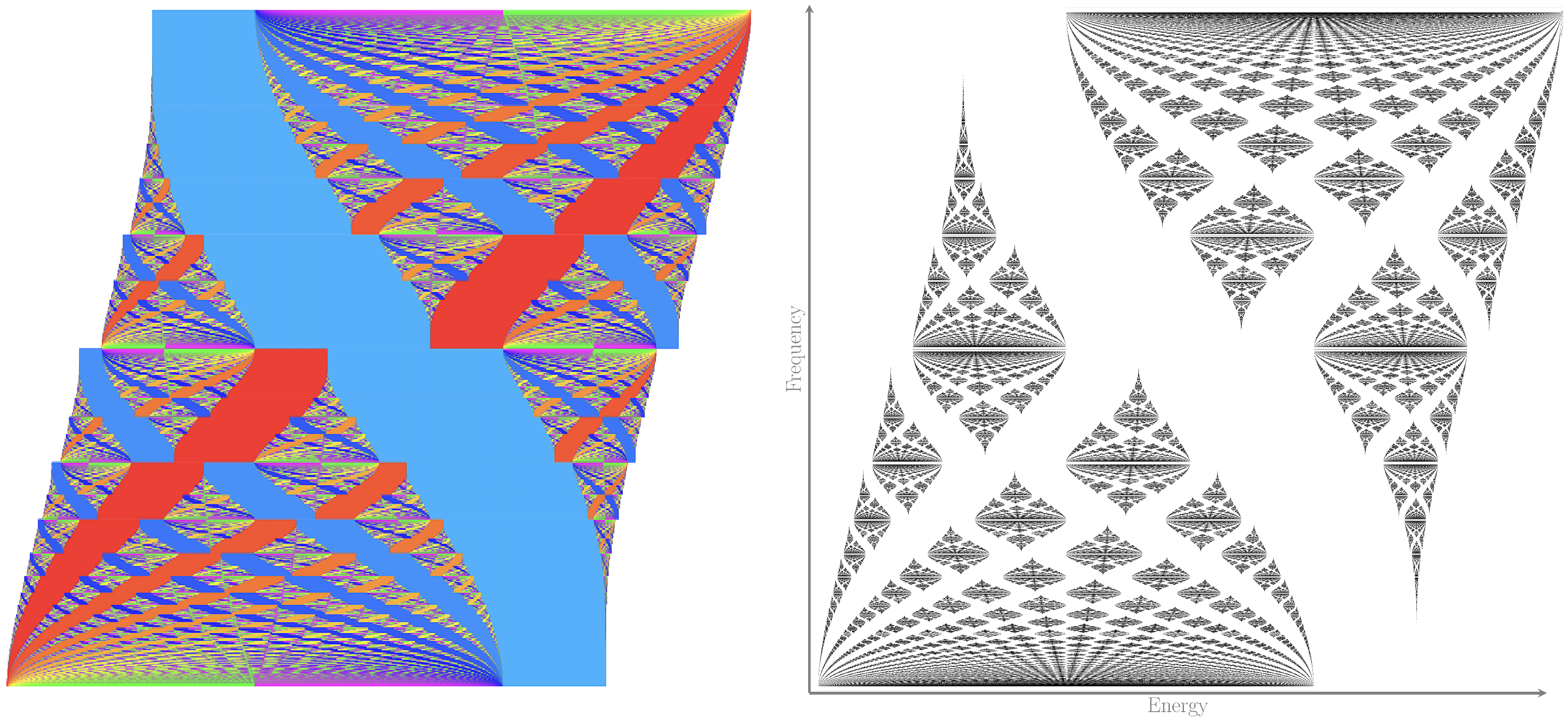}\caption{The right panel shows the Kohmoto butterfly -- the spectral bands
are plotted for different rational frequencies $\alpha$. In the left
panel each spectral gap for a periodic Hamiltonian is colored according
to its index.\protect\label{fig:=000020colored=000020butterfly}}
\end{figure}

Quasicrystals exhibit challenging spectral and topological properties.
Their quasiperiodic order gives rise to fractal spectra with infinitely
many spectral gaps, manifest in diverse wave systems \citep{OzaPriAmoGolHafLuRecSchSimZilCar_rmp2019,Cherkaev2021,Lesser2022}.

Topology together with a variety of mathematical methods enables a
classification of quasicrystals into equivalence classes governed
by topological invariants \citep{Bellissard_lecnotes_1986,ProSchBal_book2016,Kellendonk_ijc2024,DolLorSchBal_mpag2025}.
Beyond the mathematical aspects, a wide range of phenomena arise and
are studied across theoretical and experimental physics as well as
engineering \citep{Verbin2013,DarLevAguBosBouAkkGerBeu_prl17,Baboux2017,Stepanov2021Competing,Zilberberg_review2021,PhongMele2022}.
Nevertheless, there are topological shortcomings when trying to study
Kohmoto model \citep{KohKadTan_prl83}, a paradigmatic model of quasicrystals.
The three main approaches towards topological indices are not applicable
to the Kohmoto model: Chern numbers rely on differentiability of the
spectral projections \citep{ProSchBal_book2016}, and so are not defined;
the bulk--boundary correspondence (Thouless pump \citep{Thouless_PRB_1983,NiuThouless_jpa1984})
breaks down; and the two-dimensional extension (via inverse Fourier
transform) gives a parent Hamiltonian with a nonlocal slowly decaying
potential, hence not possessing a Fredholm index \citep{GraSha_cmp2018}.
This Letter overcomes these obstacles and presents an approach that
yields natural indices for the Kohmoto model. These indices consistently
reflect the quasiperiodic limit, allow to color the corresponding
phase diagram (Kohmoto butterfly) and suggest new physical invariants.

The Kohmoto model \citep{KohKadTan_prl83} is given by the Hamiltonians
\begin{align}
(\Ham\psi)(n)=\psi(n+1)+\psi(n-1)+V_{\alpha}(n)\thinspace\psi(n),\label{eq:=000020Hamiltonian=000020defined}
\end{align}
where the potential $V_{\alpha}(n)=\lambda\chi_{\left[1-\alpha,1\right)}(n\alpha\mod 1)$
is determined by a frequency $\alpha$, a coupling constant (a.k.a
modulation amplitude) $\lambda$ and $\chi_{\left[1-\alpha,1\right)}$
is the characteristic function of the interval $\left[1-\alpha,1\right)$.
It is well-known that these quasiperiodic operators represent one-dimensional
quasicrystals. For instance, the Fibonacci quasicrystal $\alpha=\frac{\sqrt{5}-1}{2}$
forms a prominent and well-studied example in this class \citep{TanGurAkk_FractEnerg14,DamGorYes_inv16}.

For rational frequencies $\alpha=\frac{p}{q}$, the operator $\Hrat$
is $q$-periodic and its spectrum consists of $q$ spectral bands.
The integrated density of states (IDS), a.k.a. electron density, is
\begin{equation}
\IDS(E)=\frac{n}{q}=\col{}\alpha\mod 1\label{eq:=000020diophantine}
\end{equation}
for energies $E$ in the $n$-th gap. The index $\col{}$ solving
the Diophantine equation above is defined only modulo $q$. The same
Diophantine equation appears in the Hofstadter model \citep{Hofstadter_prb76,Jitomirskaya_ICM2022}
of the quantum Hall effect. In that case, the modulo $q$ ambiguity
is resolved by identifying $\col{}$ with a Chern number, which can
be computed from the Berry's curvature of the corresponding spectral
projection \citep{ThoKohNigNij_prl82,DanAvrZak_jpc1985}. This endows
$\col{}$ with topological significance, as Chern numbers are invariant
under variations that do not close the spectral gap. When one assigns
a color to each integer value, this provides a consistent coloring
of the phase diagram (Hofstadter butterfly) \citep{OsaAvr_jmp01,AvrOsaSei_phystod03}.

In the Kohmoto model $\col{}$ cannot be identified with a Chern number,
since the potential $V_{\alpha}$ is discontinuous, and Berry's phase
needs the spectral projections to be differentiable \footnote{One could replace $V_{\alpha}$ by a smooth approximation, as in \citep{KelPro_ahp19}.
This is useful for irrational frequencies, but does not fully resolve
the problem for the periodic operators.}. Therefore, a coloring of the Kohmoto butterfly is not possible without
further insights for resolving the modulo $q$ ambiguity inherent
in (\ref{eq:=000020diophantine}); see \citep{AvrKenYeh_jpa14}.

We provide here a consistent coloring (depicted in Fig.~\ref{fig:=000020colored=000020butterfly},
Left) by resolving this ambiguity and determining the values of the
index invariants. We start by presenting the connection between the
periodic Hamiltonians (with $\alpha\in\Q$) and the quasiperiodic
ones ($\alpha\notin\Q$), where the former may be used to approximate
the latter.

Let $\alpha\notin\Q$ be written in terms of its continued fraction
expansion, 
\begin{equation}
\alpha=a_{0}+\frac{1}{a_{1}+\frac{1}{a_{2}+\frac{1}{\ddots}}},\label{eq:=000020infinite=000020continued=000020fraction=000020expansion}
\end{equation}
where $a_{0}=0$ and $a_{n}\in\N$ for all $n\in\N$. Truncating this
expansion gives finite continued fraction expansions, 
\begin{equation}
\alpha_{k}=a_{0}+\frac{1}{a_{1}+\frac{1}{\ddots+\frac{1}{a_{k}}}}=\frac{p_{k}}{q_{k}},\qquad k\in\N\cup\left\{ 0\right\} ,\label{eq:=000020finite=000020continued=000020fraction=000020expansion}
\end{equation}
where $p_{k},q_{k}\in\N$ are chosen to be coprime. By convention,
$\alpha_{0}=\frac{p_{0}}{q_{0}}=\frac{0}{1}$ (as $a_{0}=0$).

\begin{figure*}
\includegraphics[scale=0.9]{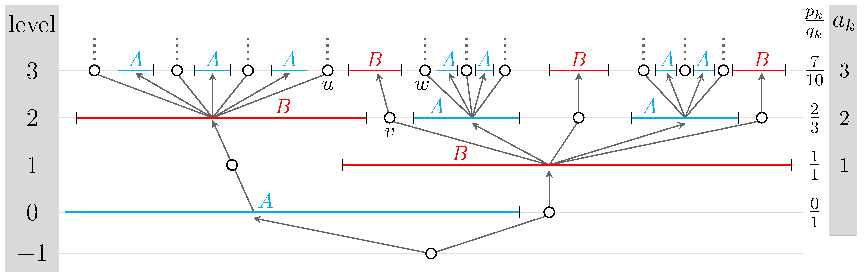}

\caption{An example of a spectral $\alpha$-tree is sketched if $\alpha$ has
continued fraction expansion $\left(a_{k}\right)_{k=0}^{\infty}$
starting with $0,1,2,3$. The vertices of the graph are drawn as the
spectral bands to which they correspond; their labels ($A/B)$ are
indicated. Vertices representing gaps $G$ are indicated by circles.
\protect\label{fig:=000020basic=000020tree=000020example}}
\end{figure*}

We construct an infinite directed tree graph $\tree$, and name it
the spectral $\alpha$-tree. This tree encodes the periodic approximations
$\Halk$ of $\Ham$. Specifically, for each $k$ the vertices at level
$k$ represent the spectral bands and gaps of $\Halk$. The tree is
constructed recursively via the digits $\{a_{1},a_{2},a_{3},\ldots\}$
of the continued fraction of $\alpha$, as illustrated in Fig.~\ref{fig:=000020basic=000020tree=000020example}
and explained next (see also \citep{BanBecLow_MFO23,BanBecLoe_arXiv24}).

We start by fixing a single vertex to be the root of the tree. We
say that the root belongs to level $k=-1$ of the tree. Starting from
the root, all other vertices belong to ascending levels $k$ in the
tree and they carry one of the three labels: $A$, $B$ or $\tC$.
The label tells whether the vertex represents a spectral gap (label
$\tC$, appearing as a circle in Fig.~\ref{fig:=000020basic=000020tree=000020example})
or a spectral band (labels $\tA$, $\tB$). The root is connected
to two vertices at level $k=0$, the left has label $\tA$ and the
right has label $\tC$. The rest of the tree $\tree$ is constructed
recursively: for every vertex $v$ with label $\tA$ or $\tB$ in
level $k\geq0$, denote
\begin{equation}
M:=\begin{cases}
a_{k+1}-1,\qquad & \text{ if }v\text{ has the label }A,\\
a_{k+1},\qquad & \text{ if }v\text{ has the label }B,
\end{cases}\label{eq:=000020branching-degree}
\end{equation}
and connect the vertex $v$ to $2M+1$ vertices in level $k+1$. The
labels of these vertices alternate between $G$ and $A$, starting
and ending with $G$, see Fig.~\ref{fig:=000020basic=000020tree=000020example}.
For a $G$-vertex $v$ in level $k$, connect it to a single $B$-vertex
in level $k+1$. This provides a complete description of $\tree$
\footnote{The description here is for $\lambda>0$. The tree for $\lambda<0$
is a vertical reflected image of the tree described above, see details
in \citep{BanBecLoe_arXiv24}.}.

Each vertex of the tree represents a spectral band or spectral gap
of $\Halk$ as shown in Fig.~\ref{fig:=000020basic=000020tree=000020example}.
The ordering of the vertices within a certain level $k$ corresponds
to the spectral order \citep{BanBecLoe_arXiv24}. In addition, if
two vertices of spectral bands (i.e., of labels $\tA/\tB$) are connected
by a directed path, it means that the upper band is fully contained
in the lower as is demonstrated in Fig.~\ref{fig:=000020basic=000020tree=000020example}.

\begin{sidewaysfigure}
\includegraphics[width=1\paperwidth]{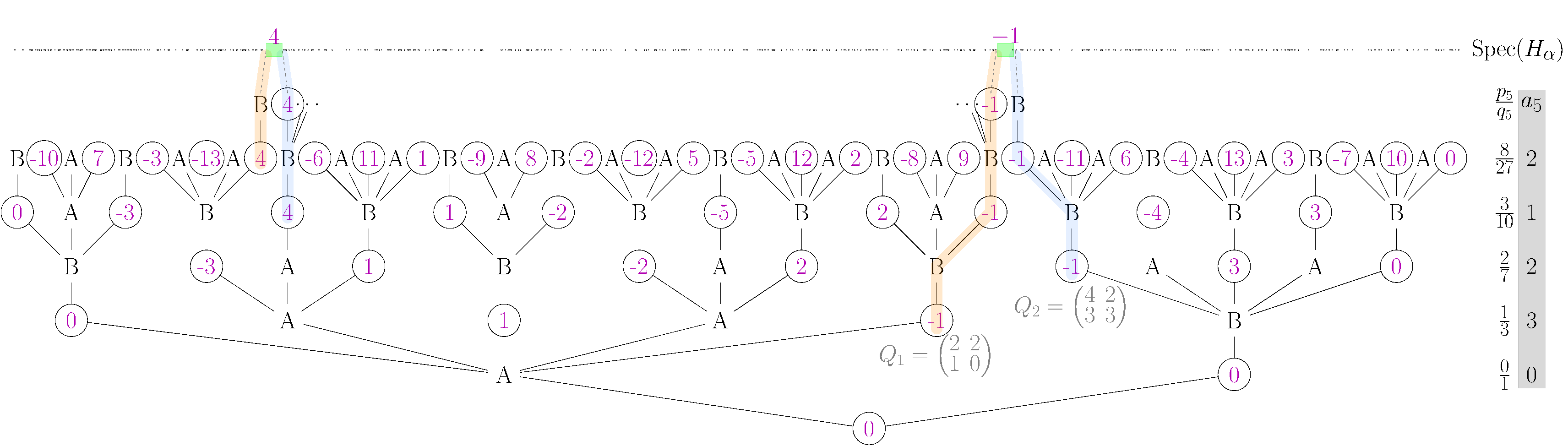}\caption{Illustration of the tree for a continued fraction beginning with $0,3,2,1,2$.
The index is marked within the circle of the $G$-vertex. Four $G$-vertices
are highlighted with the corresponding infinite paths having index
$\protect\col k(v)=-1$ respectively $\protect\col k(v)=4$.\protect\label{Fig:=000020Gap=000020index=000020-=000020Kohmoto=000020butterfly}}
\end{sidewaysfigure}

We now use the tree $\tree$ to assign indices to the spectral gaps
(i.e., to the $\tC$-vertices). For each level $k$, count the number
of $\tA$-vertices at level $k$ and denote this number by $\NA k$.
Similarly, denote the number of $\tB$-vertices in level $k$ by $\NB k$.
For each $G$-vertex $v$ in level $k$, count how many $\tA$ and
$\tB$ vertices there are at level $k$ to the left of $v$ and denote
these numbers by $\nA(v)$ and $\nB(v)$. Combine this information
to form the matrix:
\begin{equation}
\QQ k(v)=\left(\begin{array}{cc}
\NA k & \nA(v)\\
\NB k & \nB(v)
\end{array}\right).\label{eq:=000020matrix=000020Q}
\end{equation}
 See Fig.~\ref{Fig:=000020Gap=000020index=000020-=000020Kohmoto=000020butterfly}
for examples of $\QQ k(v)$ for two vertices. Using this matrix we
assign the following index to the spectral gap represented by $v$:
\begin{equation}
\col k(v)=(-1)^{k}\det\QQ k(v)\modd q_{k}.\label{eq:=000020index=000020def}
\end{equation}
The notation $\modd$ stands for the centered modulo (a.k.a. symmetric
modulo) which is defined as 
\begin{equation}
x\modd q:=x-q\left\lfloor \frac{x}{q}+\frac{1}{2}\right\rfloor \in\left[-\frac{q}{2},\frac{q}{2}\right)\cap\Z,\label{eq:=000020symmetric=000020modulu}
\end{equation}
where $\left\lfloor \cdot\right\rfloor $ is the floor function. Note
that we have slightly changed here the conventional definition of
$\modd$, by including $-\frac{q}{2}$, rather than $\frac{q}{2}$,
as is usually done. See the supplementary material, Sec.~\ref{sec:=000020Negative=000020vs=000020positive}
for an explanation of the rationale behind this choice. Fig.~\ref{Fig:=000020Gap=000020index=000020-=000020Kohmoto=000020butterfly}
demonstrates the $\col k(v)$ values which are assigned to vertices
in the few first levels of a particular tree $\tree$.

First, note that the index $\col k(v)$ in (\ref{eq:=000020index=000020def})
is indeed a solution for the Diophantine equation (\ref{eq:=000020diophantine}),
see supplementary material, Sec.~\ref{sec:=000020supp-diophantine}.
We proceed to further demonstrate that $\col k(v)$ is actually the
natural solution for the modulo $q$ ambiguity when taking into account
the governing quasiperiodic structure.

Fix a rational number $\frac{p}{q}\in\Q$. Take a spectral gap of
$\Hrat$ to which one wants to assign an index. Choose a finite continued
fraction which represents $\frac{p}{q}$ as in (\ref{eq:=000020finite=000020continued=000020fraction=000020expansion}).
Extend it arbitrarily to obtain an irrational $\alpha\not\in\Q$.
This means that there is a $k\in\N_{0}$ such that $\frac{p}{q}=\alpha_{k}$,
so $\frac{p}{q}$ is a rational approximation of $\alpha$. Now, consider
the tree $\tree$ and let $v$ be the vertex representing the chosen
spectral gap of $\Hrat$. This $G$-vertex $v$ with index $\col k(v)$
can be seen as an approximation of a particular gap of the quasiperiodic
Hamiltonian $\Ham$, as is explained below. This spectral gap of $\Ham$
has a well-defined (i.e., non-ambiguous) index $\col{}\in\Z$, such
that the IDS satisfies 
\begin{equation}
\IDS(E)=\col{}\alpha\mod 1,\quad\quad\col{}\in\Z,\label{eq:=000020IDS=000020irrational}
\end{equation}
for energies $E$ in that gap. This results from the gap labelling
theorem \citep{Bellissard1992}. Note that (\ref{eq:=000020IDS=000020irrational})
has the same form as (\ref{eq:=000020diophantine}), however, it does
not carry the same modulu ambiguity, since here $\alpha$ is irrational.
This substantial difference is a key ingredient in the ambiguity resolution.
We proceed to show that the index of the gap of $\Hrat$ coincides
with the index of the corresponding gap of $\Ham$, i.e., $\col k(v)=\col{}$.
This is independent of how we choose $\alpha$. Therefore $\col k(v)$
reflects the quasiperiodic structure and our index choice resolves
the bespoken modulo ambiguity.

To determine the mentioned spectral gap in $\Ham$, we construct (sketched
in Fig.~\ref{Fig:=000020Gap=000020index=000020-=000020Kohmoto=000020butterfly})
an infinite path $\gamma$ in the tree $\tree$ starting from $v$,
such that all the $\tC$ vertices of $\gamma$ have the index $\col{}=\col k(v)$
as well.

The construction depends on the sign of $\col k(v)$ and the parity
of $k$. Assume first that either \textbf{\emph{(i)}} $k$ is even
and $\col k(v)$ is positive, or \textbf{\emph{(ii)}} $k$ is odd
and $\col k(v)$ is negative. We set the first vertex of $\gamma$
to be $v$; the second vertex is the single $\tB$ vertex at level
$k+1$ which emanates from $v$. Afterwards in each level $k+m+1$
(for $m\geq1$), $\gamma$ is defined by choosing the right-most vertex
which emanates from its vertex at level $k+m$. This uniquely determines
a path starting at $v$ where the vertex labels alternate between
$G$ and $B$. The complementary scenario is either \textbf{\emph{(iii)}}
$k$ is even and $\col k(v)$ is negative or \textbf{\emph{(iv)}}
$k$ is odd and $\col k(v)$ is positive. Then, we act in a reverse
manner: start from $v$ and construct $\gamma$ by always picking
the left-most vertex when branching, see an example in Fig.~\ref{Fig:=000020Gap=000020index=000020-=000020Kohmoto=000020butterfly}.

We continue to verify that all $G$-vertices in $\gamma$ have the
same index. Let $v$ be a $G$-vertex at level $k$. It is connected
to a unique $B$-vertex at level $k+1$ and its neighboring $G$-vertices
are denoted by $u$ and $w$ ($u$ on the left, $w$ on the right,
as exemplified in Fig.\,\ref{fig:=000020basic=000020tree=000020example}).
The $\QQ{}$ matrices of these vertices are related as follows
\begin{align}
\QQ{k+1}(w) & =\QQ{k+1}(u)+\left(\begin{array}{cc}
0 & 0\\
0 & 1
\end{array}\right)\nonumber \\
 & =T_{k+1}\QQ k(v)+\left(\begin{array}{cc}
0 & 0\\
0 & k\mod 2
\end{array}\right),\label{eq:=000020Q-recursion}
\end{align}
where $T_{k}=\left(\begin{array}{cc}
a_{k}-1 & a_{k}\\
1 & 1
\end{array}\right)$ with $a_{k}$ being the digits of the continued fraction expansion
of $\alpha$. Validating (\ref{eq:=000020Q-recursion}) is straightforward
given the tree branching structure and that in each level we alter
between $G$-vertices and vertices with label $A$ or $B$. Now, let
$\gamma=(v_{0},v_{1},v_{2},\ldots)$ be an infinite path, as described
above, starting from $v=v_{0}$ at level $k$. Then all the even vertices
$v_{2m}$ are $G$-vertices. Using $\det T_{k}=-1$ for all $k$,
we conclude $\col k(v_{0})=\col{k+2m}(v_{2m})$ for all $m\in\N$
from (\ref{eq:=000020index=000020def}) and (\ref{eq:=000020Q-recursion}).
More details are provided in supplementary material, Sec.~\ref{sec:=000020supp-index-conservation}.

We proceed to describe the gap of $\Ham$ which corresponds to $\gamma$
and show that its index equals to the common value $\col k(v_{0})=\col{k+2m}(v_{2m})$
of all $\tC$-vertices of $\gamma$, as mentioned above. The vertices
of $\gamma$ alternate between the labels $\tC$ and $\tB$ (the first
vertex $v_{0}$ has label $\tC$). The $\tC$-vertices represent spectral
gaps of the periodic operators; these spectral gaps converge to a
spectral gap of $\Ham$, \citep{BelIocTes_cmp91}, see Sec.~\ref{sec:=000020supp-Convergence=000020Gaps}.
Furthermore, the infinite paths $\gamma$ described above come in
pairs, with each path representing one boundary (left/right) of a
spectral gap of $\Ham$ \citep{BanBecLoe_arXiv24,BanBecBibRayTho_proceedings24}, see Fig.~\ref{Fig:=000020Gap=000020index=000020-=000020Kohmoto=000020butterfly}.
The conservation law holds for all indices $\col k$ on both paths.

Pick a spectral gap $I$ of $\Ham$ and let $\gamma=(v_{0},v_{1},v_{2},\ldots)$
be the left path whose gaps (represented by the vertices $\{v_{2m}\}$)
converge to the chosen gap $I$ of $\Ham$; we choose the left path
as an example, but all arguments work as well for the right path with
suitable right/left switches. For each $m\geq0$ denote
by $E_{m}$ the right boundary of the gap represented by $v_{2m}$.
The energies $\{E_{m}\}$ converge to the right boundary of $I$ (Sec.~\ref{sec:=000020supp-Convergence=000020Gaps}),
which we denote by $E$. The IDS values of the periodic operators
at these energies satisfy $N_{\alpha_{k+2m}}(E_{m})=\col{k+2m}(v_{2m})\alk\mod 1$,
since the index $\col{k+2m}(v_{2m})$ was shown to satisfy the Diophantine
equation (\ref{eq:=000020diophantine}). By the definition of the
IDS \citep{BanBecLoe_arXiv24} together with the convergence $E_{m}\rightarrow E$
and $\alpha_{k+2m}\to\alpha$ we get $N_{\alpha_{k+2m}}(E_{m})\rightarrow\IDS(E)$
as $m\rightarrow\infty$. The gap labelling theorem \citep{Bellissard1992}
yields $\IDS(E)=\col{}\alpha\mod 1$ for some value $\col{}\in\Z$,
as in (\ref{eq:=000020IDS=000020irrational}). Hence, by the conservation
$\col k(v_{0})=\col{k+2m}(v_{2m})$ shown above and the convergence
of the IDS values, we get that this conserved index equals the index
$\col{}$ of the quasiperiodic operator $\Ham$, i.e. $\col{k+2m}(v_{2m})=\col{}$
for all $m\geq0$. This concludes the arguments justifying the Kohmoto
model indices.

To summarize, this Letter establishes a full classification of the
indices of the Kohmoto model, and while doing so three additional
goals are reached. First, the tree-based description shows that the
spectra of all periodic Kohmoto Hamiltonians are intrinsically connected,
collectively forming the Kohmoto butterfly (Fig.~\ref{fig:=000020colored=000020butterfly},
Right). Furthermore, it provides a structural framework to investigate
related quasiperiodic models and to deepen the understanding of their
indices.

Second, the tree structure exposes that the quasiperiodicity is reflected
in the periodic approximations. By the recent resolution of the dry
ten Martini problem \citep{BanBecLow_MFO23,BanBecLoe_arXiv24} it
is known that all integer values show up as an index value in (\ref{eq:=000020IDS=000020irrational}).
Explicitly, when fixing $\alpha\notin\Q$, we know that every integer
value $\col{}\in\Z$ appears as the index value of some open gap of
$\Ham$. The current Letter identifies the minimal periodic (finite-size)
approximations that realize every such possible index, and specifies
the energy gaps in which that value occurs. The ability to exactly
specify for each index in which finite-size system it appears is substantial
for experimental realizations.

Third, for the Kohmoto model we settle the ambiguity problem highlighted
in \citep{AvrKenYeh_jpa14} by confirming and sharpening the conjecture
posed there. This leads to a full coloring of the topological phase
diagram of the Kohmoto model (Fig.~\ref{fig:=000020colored=000020butterfly},
Left), in a direct analogy to Hofstadter\textquoteright s colored
butterfly~\citep{OsaAvr_jmp01,AvrOsaSei_phystod03}. Nevertheless,
the two models are substantially different. In contrary to the Hofstadter
butterfly, the complement of the Kohmoto butterfly consists of a single
connected component. Due to this, the gap indices are not restricted
to connected components of the phase diagram (Fig.~\ref{fig:=000020colored=000020butterfly})
as opposed to the Hofstadter butterfly. For example, one can see in
the colored Kohmoto butterfly that the red phases terminate without
a gap continuously shrinking and closing. This is against the folk
wisdom, grounded in smooth models. On the other hand, the color ordering
in the phase diagram is identical for both butterflies, highlighting
a similarity between the two models. This provides another perspective
on the question of topological equivalence between the Kohmoto and
Hofstadter models; a question raised in \citep{HirKoh_prl89} and
gained a substantial progress in \citep{KraZil_prl12,KelPro_ahp19},
but not yet conclusively resolved. 

\begin{acknowledgments}
We are grateful to Joseph Avron, Eric Akkermans, Johannes Kellendonk,
Laurent Raymond, Hermann Schulz-Baldes, Jacob Shapiro, Gilad Sofer,
Yannik Thomas and Oded Zilberberg for inspiring discussions.
The research for this article was partially conducted at the Israel Institute for Advanced Studies, 
as part of the Research Group Analysis, Geometry, and Spectral Theory of Graphs during 2025.
\pagebreak{}

\newpage{}
\end{acknowledgments}

\section*{Supplementary material\protect\label{part:=000020Supplementary-material}}

We add here further details about the calculations included in this
work.

\subsection{Deriving Eq.~(\ref{eq:=000020diophantine})\protect\label{sec:=000020supp-diophantine}}

Consider a rational number $\frac{p}{q}$ and an irrational $\alpha$
such that $\alk=\frac{p}{q}$. Let $v$ be a $G$-vertex representing
a spectral gap of $\spec{\Halk}$ with index $\col k(v)$. Then $v$
corresponds to the $n$-th spectral gap in $\spec{\Halk}$ with $n=\nA(v)+\nB(v)$
and $N_{\alpha_{k}}(E)=\frac{n}{q}$.

By the theory of continued fractions, we have $p_{k}q_{k-1}-p_{k-1}q_{k}=(-1)^{k-1}$,
equivalently, $p_{k}q_{k-1}=(-1)^{k-1}\mod{q_{k}}$. Since $q_{k-1}=\NA{k-1}+\NB{k-1}=\NB k$
and $\NA k+\NB k=q_{k}$ by the basic properties of the tree $\tree$,
we get $p_{k}\NB k=(-1)^{k-1}\mod{q_{k}}$ and $p_{k}\NA k=(-1)^{k}\mod{q_{k}}$.
Denoting $\ind k(v):=(-1)^{k}\det\QQ k(v)$ this implies
\begin{align*}
\ind k(v)p_{k} & \equiv(-1)^{k}\left(\NA k\nB(v)-\NB k\nA(v)\right)p_{k}\mod{q_{k}}\\
 & \equiv\nB(v)+\nA(v)\equiv n\mod{q_{k}}.
\end{align*}
By (\ref{eq:=000020index=000020def}), $\ind k(v)=\col k(v)\mod{q_{k}}$
validating the Diophantine equation (\ref{eq:=000020diophantine}).

We note that the identity $\left(p_{k}^{-1}\mod{q_{k}}\right)=(-1)^{k-1}q_{k-1}$
is a consequence of the above, and it may be used for explicitly solving
(\ref{eq:=000020diophantine}) and coloring the Kohmoto butterfly.

\subsection{Deriving Eq.~(\ref{eq:=000020Q-recursion})\protect\label{sec:=000020supp-Q-recurssion}}

Let $v$ be a $G$-vertex at level $k$ in a spectral $\alpha$-tree.
It is connected to a unique $B$-vertex $v'$ at level $k+1$, with
$u$ and $w$ the neighboring $G$-vertices. This immediately yields
\[
\QQ{k+1}(w)=\QQ{k+1}(u)+\left(\begin{array}{cc}
0 & 0\\
0 & 1
\end{array}\right).
\]
For the second equality in Eq.~(\ref{eq:=000020Q-recursion}), start
by justifying the left column, i.e. 
\begin{equation}
\left(\begin{array}{c}
\NA{k+1}\\
\NB{k+1}
\end{array}\right)=T_{k+1}\left(\begin{array}{c}
\NA k\\
\NB k
\end{array}\right).\label{eq:=000020Recursion-left}
\end{equation}
To see this note that \textbf{\emph{(i)}} all $\tA$-vertices at level
$k+1$ emanate from either $\tA$ or $\tB$ vertices at level $k$,
and their count is given by the branching degree (\ref{eq:=000020branching-degree});
and \textbf{\emph{(ii)}} there is a bijection between all $\tB$-vertices
at level $k+1$ and all $\tC$-vertices at level $k$. The number
of the latter equals the total number of both $\tA$ and $\tB$ vertices
in level $k$. It is left to show
\begin{equation}
\left(\begin{array}{c}
\nA(w)\\
\nB(w)
\end{array}\right)=T_{k+1}\left(\begin{array}{c}
\nA(v)\\
\nB(v)
\end{array}\right)+\left(\begin{array}{c}
0\\
k\mod 2
\end{array}\right).\label{eq:=000020Recursion-right}
\end{equation}
The arguments are similar to those used for (\ref{eq:=000020Recursion-left}).
The only difference is that the number of $\tC$-vertices to the left
of $v$ (and including $v$) equals the total number of $\tA$ and
$\tB$ vertices to the left of $v$, plus $(k\mod 2)$. This correction
comes since the left most vertex at odd levels is always a $\tC$-vertex.

\subsection{Index conservation\protect\label{sec:=000020supp-index-conservation}}

We establish here the index conservation along the paths $\gamma=(v_{0},v_{1},v_{2},\ldots)$,
described in the Letter. Explicitly, we show that $\col k(v_{0})=\col{k+2m}(v_{2m})$
for all $m\in\N$. As a by-product, our computations justify the choice
of the centered window $\left[-\frac{q}{2},\frac{q}{2}\right)$ in
the definition of $\col k$, (\ref{eq:=000020index=000020def}). We
start by adopting (as in Sec.~\ref{sec:=000020supp-diophantine})
the notation $\ind k(v):=(-1)^{k}\det\QQ k(v)$, with which (\ref{eq:=000020index=000020def})
reads $\col k(v)=\ind k(v)\modd q_{k}$.

Let $v$ be a $G$-vertex at level $k$ in a spectral $\alpha$-tree.
It connects to a unique $B$-vertex at level $k+1$, with $u$ and
$w$ the neighboring $G$-vertices. We begin by showing the following
identities: if $k$ is even, then 
\begin{equation}
\ind k(v)=\ind{k+1}(w)\quad\textrm{and}\quad\ind k(v)-q_{k}=\ind{k+1}(u)-q_{k+1},\label{eq:=000020ind=000020relations=000020-=000020k=000020even}
\end{equation}
and if $k$ is odd then
\begin{equation}
\ind k(v)=\ind{k+1}(u)\quad\textrm{and}\quad\ind k(v)-q_{k}=\ind{k+1}(w)-q_{k+1}.\label{eq:=000020ind=000020relations=000020-=000020k=000020odd}
\end{equation}

Since $\det(T_{k})=-1$, $\det\left(T_{k+1}Q_{k}(v)\right)=-\det\left(Q_{k}(v)\right)$.
Together with (\ref{eq:=000020Q-recursion}), this yields $\ind k(v)=\ind{k+1}(w)$
when $k$ even, and $\ind k(v)=\ind{k+1}(u)$ when $k$ is odd. This
proofs the first parts of the equations (\ref{eq:=000020ind=000020relations=000020-=000020k=000020even})
and (\ref{eq:=000020ind=000020relations=000020-=000020k=000020odd}).

For the second part, we compute
\begin{align*}
\ind k(v)-q_{k} & =(-1)^{k}\det\left[\QQ k(v)+\left(\begin{array}{cc}
0 & (-1)^{k}\\
0 & (-1)^{k+1}
\end{array}\right)\right]\\
 & =(-1)^{k+1}\det\left[T_{k+1}\left(\QQ k(v)+\left(\begin{array}{cc}
0 & (-1)^{k}\\
0 & (-1)^{k+1}
\end{array}\right)\right)\right]\\
 & =(-1)^{k+1}\det\left[T_{k+1}\QQ k(v)+\left(\begin{array}{cc}
0 & (-1)^{k+1}\\
0 & 0
\end{array}\right)\right].
\end{align*}
Thus, (\ref{eq:=000020Q-recursion}) implies if $k$ is even
\begin{align*}
\ind k(v)-q_{k} & =(-1)\det\left[\QQ{k+1}(u)+\left(\begin{array}{cc}
0 & -1\\
0 & 1
\end{array}\right)\right]\\
 & =\ind{k+1}(u)-q_{k+1}
\end{align*}
and if $k$ is odd
\begin{align*}
\ind k(v)-q_{k} & =\det\left[\QQ{k+1}(w)+\left(\begin{array}{cc}
0 & 1\\
0 & -1
\end{array}\right)\right]\\
 & =\ind{k+1}(w)-q_{k+1},
\end{align*}
which concludes the verification of (\ref{eq:=000020ind=000020relations=000020-=000020k=000020even})
and (\ref{eq:=000020ind=000020relations=000020-=000020k=000020odd})

\begin{figure}[h]
\includegraphics[scale=0.6]{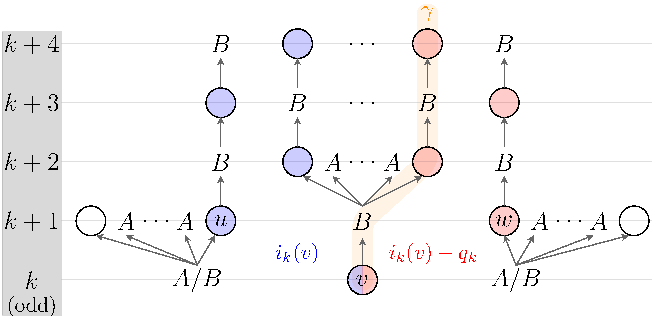}\caption{An illustration of the conservation deduced from (\ref{eq:=000020ind=000020relations=000020-=000020k=000020even})
and (\ref{eq:=000020ind=000020relations=000020-=000020k=000020odd})
for odd $k$. For all blue vertices $i_{\ast}(\ast)$ is preserved
and for all red vertices $i_{\ast}(\ast)-q_{\ast}$ is preserved.
We indicate the path $\gamma$ which is the one is chosen if $\protect\col k(v)<0$.
\protect\label{Fig:Conservation}}
\end{figure}
Equations (\ref{eq:=000020ind=000020relations=000020-=000020k=000020even})
and (\ref{eq:=000020ind=000020relations=000020-=000020k=000020odd})
show that starting from a vertex $v$, there is a path along which
$i_{k}$ is conserved and another path along which $i_{k}-q_{k}$
is conserved. The left/right orientation of those paths depends on
the parity of $k$ (see Fig.~\ref{Fig:Conservation}). Consequently,
it is natural to select the conserved quantity (either $i_{k}$ or
$i_{k}-q_{k}$) as the gap index, since this value remains invariant.

We proceed by induction to show that
\begin{equation}
i_{k}(v)\in\left[0,q_{k}\right)\label{eq:Range=000020i_k}
\end{equation}
 for all $G$-vertices $v$. For $k=0$ and $k=1$ this follows directly
from computation.

\begin{figure}[h]
\includegraphics[scale=0.6]{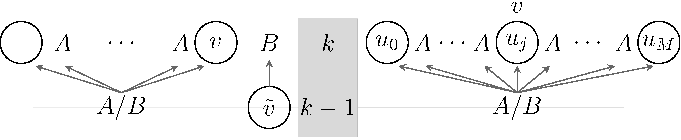}\caption{Illustration of different $G$-vertices: sandwiched between an $\protect\tA$-vertex
and a $\protect\tB$-vertex (Left) or between two $\protect\tA$-vertices
(Right). \protect\label{Fig:G-vertex=000020types}}
\end{figure}

Suppose the claim holds up to level $k-1$, and let $v$ be a $G$-vertex
in level $k$. If $v$ is a $G$-vertex with no neighbor on one side
(either left or right), then $i_{k}(v)=0$ by definition of the matrix
$Q_{k}(v)$. For all other $G$-vertices, the construction of the
spectral tree shows that either $v$ has a neighboring $B$-vertex
or both neighbors are $A$-vertices, Fig.~\ref{Fig:G-vertex=000020types}.
We treat each of these cases separately.

If, for example, there is a $B$-vertex to the right of $v$ (Fig.~\ref{Fig:G-vertex=000020types},
Left), then this $B$-vertex emanates from a $G$-vertex $\tilde{v}$
in level $k-1$. Since $i_{k}(\tilde{v})\in\left[0,q_{k}\right)$,
the relation (\ref{eq:=000020ind=000020relations=000020-=000020k=000020even})
or (\ref{eq:=000020ind=000020relations=000020-=000020k=000020odd})
implies that $\ind k(v)\in\left[0,q_{k}\right)$ using $q_{k+1}>q_{k}$.
Similarly, one shows $\ind k(v)\in\left[0,q_{k}\right)$ if there
is a $B$-vertex to the left of $v$.

Now suppose both neighbors of $v$ are $A$-vertices. Enumerate the
$G$-vertices emanating from the same $A/B$-vertex as $v$, from
left to right, by $u_{0},\ldots,u_{M}$, with $v=u_{j}$ for some
$j$ (Fig.~\ref{Fig:G-vertex=000020types}, Right). Then either there
exists a $B$-vertex to the left of $u_{0}$ (or symmetrically a $B$-vertex
to the right of $u_{M}$), or $u_{0}$ is the left-most vertex in
level $k$ (symmetrically, $u_{M}$ is the right-most vertex in level
$k$). In either case we conclude from the previous considerations
that $i_{k}(u_{0}),i_{k}(u_{N})\in\left[0,q_{k}\right)$. Since
\[
Q_{k}(u_{j})=Q_{k}(u_{0})+\begin{pmatrix}0 & j\\
0 & 0
\end{pmatrix},
\]
the sequence $i_{k}(u_{j})$ is monotone in $j$ (either decreasing
or increasing depending on the parity of $k$). Because both endpoints
$i_{k}(u_{0})$ and $i_{k}(u_{N})$ lie in $\left[0,q_{k}\right)$,
it follows that $i_{k}(u_{j})\in\left[0,q_{k}\right)$ for all $0\leq j\leq M$
and in particular for the vertex $v$. Therefore we have established
(\ref{eq:Range=000020i_k}).

By definition of $\col k$ in (\ref{eq:=000020index=000020def}) and
(\ref{eq:Range=000020i_k}) we have
\begin{equation}
\col k(v)=\begin{cases}
\ind k(v) & 0\leq\ind k(v)<\frac{q_{k}}{2},\\
\ind k(v)-q_{k} & \frac{q_{k}}{2}\leq\ind k(v)<q_{k}.
\end{cases}\label{eq:=000020index=000020def=0000202}
\end{equation}

From (\ref{eq:=000020index=000020def=0000202}) together with (\ref{eq:=000020ind=000020relations=000020-=000020k=000020even}),
(\ref{eq:=000020ind=000020relations=000020-=000020k=000020odd}),
we can now show the conservation of $\col k$ along the paths $\gamma$,
described in the Letter. We treat the case of odd $k$ (the even case
follows analogously), as illustrated in Fig.~\ref{Fig:Conservation}.
Suppose first that $\col k(v)<0$. Then the path $\gamma=(v_{0},v_{1},\ldots)$
is defined by setting $v_{0}=v$ and, at each branching, choosing
the right-most descendant. Since $\col k(v)<0$, (\ref{eq:=000020index=000020def=0000202})
gives $\col k(v)=\ind k(v)-q_{k}$. Applying (\ref{eq:=000020ind=000020relations=000020-=000020k=000020even})
and (\ref{eq:=000020ind=000020relations=000020-=000020k=000020odd})
it inductively (see the red colored vertices, arranged in a zigzag
pattern in Fig.~\ref{Fig:Conservation}) follows that $\col k(v)=\col{k+2m}(v_{2m})$
for all $m\in\N$.

Next suppose that $\col k(v)>0$. Then the path $\eta=(u_{0},u_{1},\ldots)$
is defined by setting $u_{0}=v$ and, at each branching, choosing
the left-most descendant. Since $\col k(v)>0$, (\ref{eq:=000020index=000020def=0000202})
gives $\col k(v)=\ind k(v)$. Applying (\ref{eq:=000020ind=000020relations=000020-=000020k=000020even})
and (\ref{eq:=000020ind=000020relations=000020-=000020k=000020odd})
it inductively (see the blue colored vertices, arranged in a zigzag
pattern in Fig.~\ref{Fig:Conservation}) follows that $\col k(v)=\col{k+2m}(u_{2m})$
for all $m\in\N$.

\subsection{Spectral gaps convergence along $\gamma$\protect\label{sec:=000020supp-Convergence=000020Gaps}}

Consider an irrational $\alpha$ with its spectral $\alpha$-tree,
$\tree$. Let $v$ be a $G$-vertex with index $\col k(v)$ and let
$\gamma=(v_{0},v_{1},\dots)$ be the infinite path starting at $v=v_{0}$,
which is defined in this Letter, such that the indices along it are
conserved, i.e., $\col k(v)=\col{k+2m}(v_{2m})$ for all $m\in\N$.
For each $v_{2m}$, the open interval $I_{m}=(L_{m},R_{m})$ denotes
the spectral gap of $\spec{H_{\alpha_{k+2m}}}$ represented by~$v_{2m}$.
Our goal is to verify that these spectral gaps $I_{m}$ converge to
the limiting spectral gap $I=(L,R)$ in $\spec{\Ham}$, which carries
the same index $\col k(v)$.

We have two cases: either $v_{2m}$ is the right-most vertex emanating
from $v_{2m-1}$ for all $m\in\N$ or it is always the left-most vertex.
Without loss of generality we assume that $v_{2m}$ is the right-most
vertex for all $m\in\N$. We note that there is a neighboring path
$\tilde{\gamma}=(w_{0},w_{1},\ldots)$ whose $G$-vertices share the
same index values as the $G$-vertices of $\gamma$. For this path,
$w_{0}$ is set to be the neighboring vertex at level $k+1$ to the
right of the $B$-vertex emanating from $v_{0}$ (see Fig.~\ref{Fig:Conservation}
with $v_{0}=v$ and $w_{0}=w$). For the rest of the vertices of $\tilde{\gamma}$
we choose $w_{2m}$ to be the left-most vertex emanating from $w_{2m-1}$
for all $m\in\N$. By this construction we get that the vertex $w_{m}$
is the right neighbor of the vertex $v_{m+1}$ for all $m\in\N$.

Recall that the spectral gap associated with the $\tC$-vertex $v_{2m}$
is $I_{m}=(L_{m},R_{m})$. By construction, the right endpoint $R_{m}\in\spec{H_{\alpha_{k+2m}}}$
belongs to the spectral band associated with the $B$-vertex $w_{2m+1}$.
These $B$-vertices correspond to spectral bands of the periodic operators,
which form a decreasing nested sequence. Their intersection is a single
point \citep{BanBecLoe_arXiv24}, lying in $\spec{\Ham}$ and serving
as the right endpoint $R$ of the limiting gap $I$. Thus $R_{m}\to R$
as $m\to\infty$.

Moreover, $\spec{H_{\alpha_{k+2m}}}\to\spec{\Ham}$ as $m\to\infty$
\citep{BelIocTes_cmp91}, which implies that the left endpoints also
converge, $L_{m}\to L$. Hence the gaps $I_{m}=(L_{m},R_{m})$ converge
to the limiting gap $I=(L,R)$ of $\spec{\Ham}$.

\subsection{Negative indices versus positive indices\protect\label{sec:=000020Negative=000020vs=000020positive}}

We comment here on the comparison between negative values of $\col k(v)$
versus positive values. For this discussion, recall the notation $\ind k(v):=(-1)^{k}\det\QQ k(v)$
(as in Sec.~\ref{sec:=000020supp-diophantine} and \ref{sec:=000020supp-index-conservation})
and the connection (\ref{eq:=000020index=000020def=0000202}) between
$i_{k}(v)$ and $\col k(v)$. We discuss to the particular case $i_{k}(v)=\frac{q_{k}}{2}$
for which one should determine the sign for $\col k(v)=\pm\frac{q_{k}}{2}$.
This decision breaks the symmetry of the modulus window chosen for
$\modd$, i.e., whether the image of the modulus is $\left[-\frac{q_{k}}{2},\frac{q_{k}}{2}\right)$
or $\left(-\frac{q_{k}}{2},\frac{q_{k}}{2}\right]$. This decision
was already made in (\ref{eq:=000020index=000020def}) (see also (\ref{eq:=000020index=000020def=0000202}))
and we justify it here.

\begin{figure}[h]
\includegraphics[scale=0.6]{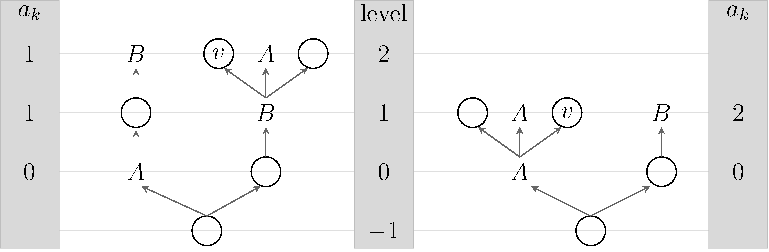}\caption{Two examples of spectral trees with a marked vertex $v$ in level
$k$ which is either sandwiched between $\protect\tB$ and $\protect\tA$
vertices and $k$ is even (Left) or $v$ is sandwiched between $\protect\tA$
and $\protect\tB$ vertices and $k$ is odd (Right). \protect\label{Fig:=000020negative=000020index}}
\end{figure}
As a guiding example we take $q_{k}=2$. In this case, there are two
spectral bands and only one bounded gap. We discuss the possible index
value of that gap. To do so, consider a spectral tree $\tree$ with
$\alpha$ having a continued fraction expansion starting with $2$
(see Fig.~\ref{Fig:=000020negative=000020index},\,Right). In this
tree there is a vertex $v$ in level $k=1$ which corresponds to the
bounded gap we referred to, and indeed $q_{1}=2$. For this vertex
$i_{1}(v)=1=\frac{q_{1}}{2}$, and we wish to explain the choice made
in (\ref{eq:=000020index=000020def}) for the modulus which gives
$\col 1(v)=-1$. Alternatively, $q_{k}=2$ can be obtained from another
spectral tree, $\tree$, where $\alpha$ has the continued fraction
digits $1,1,\ldots$ (see Fig.~\ref{Fig:=000020negative=000020index},\,Left).
In this case there is a vertex $v$ in level $k=2$ for which $i_{2}(v)=1=\frac{q_{2}}{2}$.

In the first case above the vertex $v$ is sandwiched between $\tA$
and $\tB$ vertices (in that order) and $k$ is odd. In the second
case, the vertex $v$ is sandwiched between $\tB$ and $\tA$ vertices
and $k$ is even. These two cases belong to the same general class
for all spectral trees: if either \textbf{\emph{(i)}} $v$ has $\tA$-vertex
to its left and $\tB$-vertex to its right and $k$ is odd or \textbf{\emph{(ii)}}
$v$ has $\tB$-vertex to its left and $\tA$-vertex to its right
and $k$ is even, then $i_{k}(v)\geq\frac{q_{k}}{2}$. This justifies
that the value $i_{k}(v)=\frac{q_{k}}{2}$ behaves under the modulus
operation similarly to the values $i_{k}(v)\in(\frac{q_{k}}{2},q_{k})$,
see (\ref{eq:=000020index=000020def=0000202}), and hence a negative
value for the index is obtained, $\col k(v)=-\frac{q_{k}}{2}$. We
mention also the counterpart behavior: if either \textbf{\emph{(iii)}}
$v$ has $\tA$-vertex to its left and $\tB$-vertex to its right
and $k$ is even or \textbf{\emph{(iv)}} $v$ has $\tB$-vertex to
its left and $\tA$-vertex to its right and $k$ is odd, then $i_{k}(v)<\frac{q_{k}}{2}$
(and $\col k(v)$ gets a positive value). The general statement above
(with all of its parts (i)-(iv)) can be shown by induction, but we
omit here the technical proof, and merely refer to Fig.\,\ref{Fig:=000020Gap=000020index=000020-=000020Kohmoto=000020butterfly},
which exemplifies it.

We complement this discussion with an additional viewpoint on negative
versus positive index values for $\col k(v)$. One observes (Fig.~\ref{fig:=000020colored=000020butterfly},
Left) that the larger the absolute value of the index is, the smaller
is the spectral gap and if the absolute value of two indices agree,
then the one with negative index is dominant \footnote{This interesting observation yet awaits a rigorous explanation.}.
In particular, the gaps which correspond to the $\tC$-vertices whose
index equals $-1$ are wider than those of index value $1$ for all
values of $q_{k}$, and hence more dominant and preferable in terms
of index choice.

\sloppy
\bibliography{GlobalBib_250731,quasicrystal_merged_refs}

\begin{thebibliography}{37}%
\makeatletter
\providecommand \@ifxundefined [1]{%
 \@ifx{#1\undefined}
}%
\providecommand \@ifnum [1]{%
 \ifnum #1\expandafter \@firstoftwo
 \else \expandafter \@secondoftwo
 \fi
}%
\providecommand \@ifx [1]{%
 \ifx #1\expandafter \@firstoftwo
 \else \expandafter \@secondoftwo
 \fi
}%
\providecommand \natexlab [1]{#1}%
\providecommand \enquote  [1]{``#1''}%
\providecommand \bibnamefont  [1]{#1}%
\providecommand \bibfnamefont [1]{#1}%
\providecommand \citenamefont [1]{#1}%
\providecommand \href@noop [0]{\@secondoftwo}%
\providecommand \href [0]{\begingroup \@sanitize@url \@href}%
\providecommand \@href[1]{\@@startlink{#1}\@@href}%
\providecommand \@@href[1]{\endgroup#1\@@endlink}%
\providecommand \@sanitize@url [0]{\catcode `\\12\catcode `\$12\catcode
  `\&12\catcode `\#12\catcode `\^12\catcode `\_12\catcode `\%12\relax}%
\providecommand \@@startlink[1]{}%
\providecommand \@@endlink[0]{}%
\providecommand \url  [0]{\begingroup\@sanitize@url \@url }%
\providecommand \@url [1]{\endgroup\@href {#1}{\urlprefix }}%
\providecommand \urlprefix  [0]{URL }%
\providecommand \Eprint [0]{\href }%
\providecommand \doibase [0]{https://doi.org/}%
\providecommand \selectlanguage [0]{\@gobble}%
\providecommand \bibinfo  [0]{\@secondoftwo}%
\providecommand \bibfield  [0]{\@secondoftwo}%
\providecommand \translation [1]{[#1]}%
\providecommand \BibitemOpen [0]{}%
\providecommand \bibitemStop [0]{}%
\providecommand \bibitemNoStop [0]{.\EOS\space}%
\providecommand \EOS [0]{\spacefactor3000\relax}%
\providecommand \BibitemShut  [1]{\csname bibitem#1\endcsname}%
\let\auto@bib@innerbib\@empty
\bibitem [{\citenamefont {Ozawa}\ \emph {et~al.}(2019)\citenamefont {Ozawa},
  \citenamefont {Price}, \citenamefont {Amo}, \citenamefont {Goldman},
  \citenamefont {Hafezi}, \citenamefont {Lu}, \citenamefont {Rechtsman},
  \citenamefont {Schuster}, \citenamefont {Simon}, \citenamefont {Zilberberg},\
  and\ \citenamefont {Carusotto}}]{OzaPriAmoGolHafLuRecSchSimZilCar_rmp2019}%
  \BibitemOpen
  \bibfield  {author} {\bibinfo {author} {\bibfnamefont {T.}~\bibnamefont
  {Ozawa}}, \bibinfo {author} {\bibfnamefont {H.~M.}\ \bibnamefont {Price}},
  \bibinfo {author} {\bibfnamefont {A.}~\bibnamefont {Amo}}, \bibinfo {author}
  {\bibfnamefont {N.}~\bibnamefont {Goldman}}, \bibinfo {author} {\bibfnamefont
  {M.}~\bibnamefont {Hafezi}}, \bibinfo {author} {\bibfnamefont
  {L.}~\bibnamefont {Lu}}, \bibinfo {author} {\bibfnamefont {M.~C.}\
  \bibnamefont {Rechtsman}}, \bibinfo {author} {\bibfnamefont {D.}~\bibnamefont
  {Schuster}}, \bibinfo {author} {\bibfnamefont {J.}~\bibnamefont {Simon}},
  \bibinfo {author} {\bibfnamefont {O.}~\bibnamefont {Zilberberg}},\ and\
  \bibinfo {author} {\bibfnamefont {I.}~\bibnamefont {Carusotto}},\ }\bibfield
  {title} {\bibinfo {title} {Topological photonics},\ }\href
  {https://doi.org/10.1103/RevModPhys.91.015006} {\bibfield  {journal}
  {\bibinfo  {journal} {Reviews of Modern Physics}\ }\textbf {\bibinfo {volume}
  {91}},\ \bibinfo {pages} {015006} (\bibinfo {year} {2019})}\BibitemShut
  {NoStop}%
\bibitem [{\citenamefont {Cherkaev}\ \emph {et~al.}(2021)\citenamefont
  {Cherkaev}, \citenamefont {Vasquez}, \citenamefont {Mauck}, \citenamefont
  {Prisbrey},\ and\ \citenamefont {Raeymaekers}}]{Cherkaev2021}%
  \BibitemOpen
  \bibfield  {author} {\bibinfo {author} {\bibfnamefont {E.}~\bibnamefont
  {Cherkaev}}, \bibinfo {author} {\bibfnamefont {F.~G.}\ \bibnamefont
  {Vasquez}}, \bibinfo {author} {\bibfnamefont {C.}~\bibnamefont {Mauck}},
  \bibinfo {author} {\bibfnamefont {M.}~\bibnamefont {Prisbrey}},\ and\
  \bibinfo {author} {\bibfnamefont {B.}~\bibnamefont {Raeymaekers}},\
  }\bibfield  {title} {\bibinfo {title} {Wave-driven assembly of quasiperiodic
  patterns of particles},\ }\href
  {https://doi.org/10.1103/PhysRevLett.126.145501} {\bibfield  {journal}
  {\bibinfo  {journal} {Phys. Rev. Lett.}\ }\textbf {\bibinfo {volume} {126}},\
  \bibinfo {pages} {145501} (\bibinfo {year} {2021})}\BibitemShut {NoStop}%
\bibitem [{\citenamefont {Lesser}\ and\ \citenamefont
  {Lifshitz}(2022)}]{Lesser2022}%
  \BibitemOpen
  \bibfield  {author} {\bibinfo {author} {\bibfnamefont {O.}~\bibnamefont
  {Lesser}}\ and\ \bibinfo {author} {\bibfnamefont {R.}~\bibnamefont
  {Lifshitz}},\ }\bibfield  {title} {\bibinfo {title} {Emergence of
  quasiperiodic {B}loch wave functions in quasicrystals},\ }\href
  {https://doi.org/10.1103/PhysRevResearch.4.013226} {\bibfield  {journal}
  {\bibinfo  {journal} {Phys. Rev. Research}\ }\textbf {\bibinfo {volume}
  {4}},\ \bibinfo {pages} {013226} (\bibinfo {year} {2022})}\BibitemShut
  {NoStop}%
\bibitem [{\citenamefont {Bellissard}(1986)}]{Bellissard_lecnotes_1986}%
  \BibitemOpen
  \bibfield  {author} {\bibinfo {author} {\bibfnamefont {J.}~\bibnamefont
  {Bellissard}},\ }\bibfield  {title} {\bibinfo {title} {{K}-theory of
  {C}*-algebras in solid state physics},\ }in\ \href
  {https://doi.org/10.1007/3-540-16777-3\_74} {\emph {\bibinfo {booktitle}
  {Statistical Mechanics and Field Theory: Mathematical Aspects}}},\ \bibinfo
  {series} {Lecture Notes in Physics}, Vol.\ \bibinfo {volume} {257},\ \bibinfo
  {editor} {edited by\ \bibinfo {editor} {\bibfnamefont {T.~C.}\ \bibnamefont
  {Dorlas}}, \bibinfo {editor} {\bibfnamefont {N.~M.}\ \bibnamefont
  {Hugenholtz}},\ and\ \bibinfo {editor} {\bibfnamefont {M.}~\bibnamefont
  {Winnink}}}\ (\bibinfo  {publisher} {Springer, Berlin, Heidelberg},\ \bibinfo
  {year} {1986})\ pp.\ \bibinfo {pages} {99--156}\BibitemShut {NoStop}%
\bibitem [{\citenamefont {Prodan}\ and\ \citenamefont
  {Schulz-Baldes}(2016)}]{ProSchBal_book2016}%
  \BibitemOpen
  \bibfield  {author} {\bibinfo {author} {\bibfnamefont {E.}~\bibnamefont
  {Prodan}}\ and\ \bibinfo {author} {\bibfnamefont {H.}~\bibnamefont
  {Schulz-Baldes}},\ }\href@noop {} {\emph {\bibinfo {title} {Bulk and Boundary
  Invariants for Complex Topological \allowbreak\ Insulators: From {K}-{T}heory
  to Physics}}},\ Mathematical Physics Studies\ (\bibinfo  {publisher}
  {Springer International Publishing, Cham},\ \bibinfo {year}
  {2016})\BibitemShut {NoStop}%
\bibitem [{\citenamefont {Kellendonk}(2024)}]{Kellendonk_ijc2024}%
  \BibitemOpen
  \bibfield  {author} {\bibinfo {author} {\bibfnamefont {J.}~\bibnamefont
  {Kellendonk}},\ }\bibfield  {title} {\bibinfo {title} {Topological quantum
  numbers in quasicrystals},\ }\bibfield  {journal} {\bibinfo  {journal}
  {Israel Journal of Chemistry}\ }\href
  {https://doi.org/10.1002/ijch.202400027} {10.1002/ijch.202400027} (\bibinfo
  {year} {2024}),\ \bibinfo {note} {overview on the theory of topological
  quantum numbers in quasicrystals via non-commutative topology}\BibitemShut
  {NoStop}%
\bibitem [{\citenamefont {Doll}\ \emph {et~al.}(2025)\citenamefont {Doll},
  \citenamefont {Loring},\ and\ \citenamefont
  {Schulz-Baldes}}]{DolLorSchBal_mpag2025}%
  \BibitemOpen
  \bibfield  {author} {\bibinfo {author} {\bibfnamefont {N.}~\bibnamefont
  {Doll}}, \bibinfo {author} {\bibfnamefont {T.}~\bibnamefont {Loring}},\ and\
  \bibinfo {author} {\bibfnamefont {H.}~\bibnamefont {Schulz-Baldes}},\
  }\bibfield  {title} {\bibinfo {title} {Topological indices for periodic
  gapped hamiltonians and fuzzy tori},\ }\bibfield  {journal} {\bibinfo
  {journal} {Mathematical Physics, Analysis and Geometry}\ }\textbf {\bibinfo
  {volume} {28}},\ \href {https://doi.org/10.1007/s11040-025-09508-0}
  {10.1007/s11040-025-09508-0} (\bibinfo {year} {2025})\BibitemShut {NoStop}%
\bibitem [{\citenamefont {Verbin}\ \emph {et~al.}(2013)\citenamefont {Verbin},
  \citenamefont {Zilberberg}, \citenamefont {Kraus}, \citenamefont {Lahini},\
  and\ \citenamefont {Silberberg}}]{Verbin2013}%
  \BibitemOpen
  \bibfield  {author} {\bibinfo {author} {\bibfnamefont {M.}~\bibnamefont
  {Verbin}}, \bibinfo {author} {\bibfnamefont {O.}~\bibnamefont {Zilberberg}},
  \bibinfo {author} {\bibfnamefont {Y.~E.}\ \bibnamefont {Kraus}}, \bibinfo
  {author} {\bibfnamefont {Y.}~\bibnamefont {Lahini}},\ and\ \bibinfo {author}
  {\bibfnamefont {Y.}~\bibnamefont {Silberberg}},\ }\bibfield  {title}
  {\bibinfo {title} {Observation of topological phase transitions in photonic
  quasicrystals},\ }\href {https://doi.org/10.1103/PhysRevLett.110.076403}
  {\bibfield  {journal} {\bibinfo  {journal} {Phys. Rev. Lett.}\ }\textbf
  {\bibinfo {volume} {110}},\ \bibinfo {pages} {076403} (\bibinfo {year}
  {2013})}\BibitemShut {NoStop}%
\bibitem [{\citenamefont {Dareau}\ \emph {et~al.}(2017)\citenamefont {Dareau},
  \citenamefont {Levy}, \citenamefont {Aguilera}, \citenamefont {Bouganne},
  \citenamefont {Akkermans}, \citenamefont {Gerbier},\ and\ \citenamefont
  {Beugnon}}]{DarLevAguBosBouAkkGerBeu_prl17}%
  \BibitemOpen
  \bibfield  {author} {\bibinfo {author} {\bibfnamefont {A.}~\bibnamefont
  {Dareau}}, \bibinfo {author} {\bibfnamefont {E.}~\bibnamefont {Levy}},
  \bibinfo {author} {\bibfnamefont {M.~B.}\ \bibnamefont {Aguilera}}, \bibinfo
  {author} {\bibfnamefont {R.}~\bibnamefont {Bouganne}}, \bibinfo {author}
  {\bibfnamefont {E.}~\bibnamefont {Akkermans}}, \bibinfo {author}
  {\bibfnamefont {F.}~\bibnamefont {Gerbier}},\ and\ \bibinfo {author}
  {\bibfnamefont {J.}~\bibnamefont {Beugnon}},\ }\bibfield  {title} {\bibinfo
  {title} {Revealing the topology of quasicrystals with a diffraction
  experiment},\ }\href {https://doi.org/10.1103/PhysRevLett.119.215304}
  {\bibfield  {journal} {\bibinfo  {journal} {Phys. Rev. Lett.}\ }\textbf
  {\bibinfo {volume} {119}},\ \bibinfo {pages} {215304} (\bibinfo {year}
  {2017})}\BibitemShut {NoStop}%
\bibitem [{\citenamefont {Baboux}\ \emph {et~al.}(2017)\citenamefont {Baboux},
  \citenamefont {Levy}, \citenamefont {Lema\^itre}, \citenamefont {G\'omez},
  \citenamefont {Galopin}, \citenamefont {Gratiet}, \citenamefont {Sagnes},
  \citenamefont {Amo}, \citenamefont {Bloch},\ and\ \citenamefont
  {Akkermans}}]{Baboux2017}%
  \BibitemOpen
  \bibfield  {author} {\bibinfo {author} {\bibfnamefont {F.}~\bibnamefont
  {Baboux}}, \bibinfo {author} {\bibfnamefont {E.}~\bibnamefont {Levy}},
  \bibinfo {author} {\bibfnamefont {A.}~\bibnamefont {Lema\^itre}}, \bibinfo
  {author} {\bibfnamefont {C.}~\bibnamefont {G\'omez}}, \bibinfo {author}
  {\bibfnamefont {E.}~\bibnamefont {Galopin}}, \bibinfo {author} {\bibfnamefont
  {L.~L.}\ \bibnamefont {Gratiet}}, \bibinfo {author} {\bibfnamefont
  {I.}~\bibnamefont {Sagnes}}, \bibinfo {author} {\bibfnamefont
  {A.}~\bibnamefont {Amo}}, \bibinfo {author} {\bibfnamefont {J.}~\bibnamefont
  {Bloch}},\ and\ \bibinfo {author} {\bibfnamefont {E.}~\bibnamefont
  {Akkermans}},\ }\bibfield  {title} {\bibinfo {title} {Measuring topological
  invariants from generalized edge states in polaritonic quasicrystals},\
  }\href {https://doi.org/10.1103/PhysRevB.95.161114} {\bibfield  {journal}
  {\bibinfo  {journal} {Phys. Rev. B}\ }\textbf {\bibinfo {volume} {95}},\
  \bibinfo {pages} {161114} (\bibinfo {year} {2017})}\BibitemShut {NoStop}%
\bibitem [{\citenamefont {Stepanov}\ \emph {et~al.}(2021)\citenamefont
  {Stepanov}, \citenamefont {Xie}, \citenamefont {Taniguchi}, \citenamefont
  {Watanabe}, \citenamefont {Lu}, \citenamefont {MacDonald}, \citenamefont
  {Bernevig},\ and\ \citenamefont {Efetov}}]{Stepanov2021Competing}%
  \BibitemOpen
  \bibfield  {author} {\bibinfo {author} {\bibfnamefont {P.}~\bibnamefont
  {Stepanov}}, \bibinfo {author} {\bibfnamefont {M.}~\bibnamefont {Xie}},
  \bibinfo {author} {\bibfnamefont {T.}~\bibnamefont {Taniguchi}}, \bibinfo
  {author} {\bibfnamefont {K.}~\bibnamefont {Watanabe}}, \bibinfo {author}
  {\bibfnamefont {X.}~\bibnamefont {Lu}}, \bibinfo {author} {\bibfnamefont
  {A.~H.}\ \bibnamefont {MacDonald}}, \bibinfo {author} {\bibfnamefont {B.~A.}\
  \bibnamefont {Bernevig}},\ and\ \bibinfo {author} {\bibfnamefont {D.~K.}\
  \bibnamefont {Efetov}},\ }\bibfield  {title} {\bibinfo {title} {Competing
  zero-field chern insulators in superconducting twisted bilayer graphene},\
  }\href {https://doi.org/10.1103/PhysRevLett.127.197701} {\bibfield  {journal}
  {\bibinfo  {journal} {Physical Review Letters}\ }\textbf {\bibinfo {volume}
  {127}},\ \bibinfo {pages} {197701} (\bibinfo {year} {2021})}\BibitemShut
  {NoStop}%
\bibitem [{\citenamefont {Zilberberg}(2021)}]{Zilberberg_review2021}%
  \BibitemOpen
  \bibfield  {author} {\bibinfo {author} {\bibfnamefont {O.}~\bibnamefont
  {Zilberberg}},\ }\bibfield  {title} {\bibinfo {title} {Topology in
  quasicrystals},\ }\href {https://doi.org/10.1364/OME.416552} {\bibfield
  {journal} {\bibinfo  {journal} {Optical Materials Express}\ }\textbf
  {\bibinfo {volume} {11}},\ \bibinfo {pages} {1143} (\bibinfo {year}
  {2021})}\BibitemShut {NoStop}%
\bibitem [{\citenamefont {Phong}\ and\ \citenamefont
  {Mele}(2022)}]{PhongMele2022}%
  \BibitemOpen
  \bibfield  {author} {\bibinfo {author} {\bibfnamefont {V.~T.}\ \bibnamefont
  {Phong}}\ and\ \bibinfo {author} {\bibfnamefont {E.~J.}\ \bibnamefont
  {Mele}},\ }\bibfield  {title} {\bibinfo {title} {Boundary modes from periodic
  magnetic and pseudomagnetic fields in graphene},\ }\href
  {https://doi.org/10.1103/PhysRevLett.128.176406} {\bibfield  {journal}
  {\bibinfo  {journal} {Phys. Rev. Lett.}\ }\textbf {\bibinfo {volume} {128}},\
  \bibinfo {pages} {176406} (\bibinfo {year} {2022})}\BibitemShut {NoStop}%
\bibitem [{\citenamefont {Kohmoto}\ \emph {et~al.}(1983)\citenamefont
  {Kohmoto}, \citenamefont {Kadanoff},\ and\ \citenamefont
  {Tang}}]{KohKadTan_prl83}%
  \BibitemOpen
  \bibfield  {author} {\bibinfo {author} {\bibfnamefont {M.}~\bibnamefont
  {Kohmoto}}, \bibinfo {author} {\bibfnamefont {L.~P.}\ \bibnamefont
  {Kadanoff}},\ and\ \bibinfo {author} {\bibfnamefont {C.}~\bibnamefont
  {Tang}},\ }\bibfield  {title} {\bibinfo {title} {Localization problem in one
  dimension: Mapping and escape},\ }\href
  {https://doi.org/10.1103/PhysRevLett.50.1870} {\bibfield  {journal} {\bibinfo
   {journal} {Phys. Rev. Lett.}\ }\textbf {\bibinfo {volume} {50}},\ \bibinfo
  {pages} {1870} (\bibinfo {year} {1983})}\BibitemShut {NoStop}%
\bibitem [{\citenamefont {Thouless}(1983)}]{Thouless_PRB_1983}%
  \BibitemOpen
  \bibfield  {author} {\bibinfo {author} {\bibfnamefont {D.~J.}\ \bibnamefont
  {Thouless}},\ }\bibfield  {title} {\bibinfo {title} {Quantization of particle
  transport},\ }\href {https://doi.org/10.1103/PhysRevB.27.6083} {\bibfield
  {journal} {\bibinfo  {journal} {Physical Review B}\ }\textbf {\bibinfo
  {volume} {27}},\ \bibinfo {pages} {6083} (\bibinfo {year}
  {1983})}\BibitemShut {NoStop}%
\bibitem [{\citenamefont {Niu}\ and\ \citenamefont
  {Thouless}(1984)}]{NiuThouless_jpa1984}%
  \BibitemOpen
  \bibfield  {author} {\bibinfo {author} {\bibfnamefont {Q.}~\bibnamefont
  {Niu}}\ and\ \bibinfo {author} {\bibfnamefont {D.~J.}\ \bibnamefont
  {Thouless}},\ }\bibfield  {title} {\bibinfo {title} {Quantised adiabatic
  charge transport in the presence of substrate disorder and many-body
  interaction},\ }\bibfield  {journal} {\bibinfo  {journal} {Journal of Physics
  A: Mathematical and General}\ }\textbf {\bibinfo {volume} {17}},\ \href
  {https://doi.org/10.1088/0305-4470/17/12/016} {10.1088/0305-4470/17/12/016}
  (\bibinfo {year} {1984})\BibitemShut {NoStop}%
\bibitem [{\citenamefont {Graf}\ and\ \citenamefont
  {Shapiro}(2018)}]{GraSha_cmp2018}%
  \BibitemOpen
  \bibfield  {author} {\bibinfo {author} {\bibfnamefont {G.~M.}\ \bibnamefont
  {Graf}}\ and\ \bibinfo {author} {\bibfnamefont {J.}~\bibnamefont {Shapiro}},\
  }\bibfield  {title} {\bibinfo {title} {The bulk--edge correspondence for
  disordered chiral chains},\ }\href
  {https://doi.org/10.1007/s00220-018-3247-0} {\bibfield  {journal} {\bibinfo
  {journal} {Communications in Mathematical Physics}\ }\textbf {\bibinfo
  {volume} {363}},\ \bibinfo {pages} {829} (\bibinfo {year}
  {2018})}\BibitemShut {NoStop}%
\bibitem [{\citenamefont {Tanese}\ \emph {et~al.}(2014)\citenamefont {Tanese},
  \citenamefont {Gurevich}, \citenamefont {Baboux}, \citenamefont {Jacqmin},
  \citenamefont {Lema\^{\i}tre}, \citenamefont {Galopin}, \citenamefont
  {Sagnes}, \citenamefont {Amo}, \citenamefont {Bloch},\ and\ \citenamefont
  {Akkermans}}]{TanGurAkk_FractEnerg14}%
  \BibitemOpen
  \bibfield  {author} {\bibinfo {author} {\bibfnamefont {D.}~\bibnamefont
  {Tanese}}, \bibinfo {author} {\bibfnamefont {E.}~\bibnamefont {Gurevich}},
  \bibinfo {author} {\bibfnamefont {F.}~\bibnamefont {Baboux}}, \bibinfo
  {author} {\bibfnamefont {T.}~\bibnamefont {Jacqmin}}, \bibinfo {author}
  {\bibfnamefont {A.}~\bibnamefont {Lema\^{\i}tre}}, \bibinfo {author}
  {\bibfnamefont {E.}~\bibnamefont {Galopin}}, \bibinfo {author} {\bibfnamefont
  {I.}~\bibnamefont {Sagnes}}, \bibinfo {author} {\bibfnamefont
  {A.}~\bibnamefont {Amo}}, \bibinfo {author} {\bibfnamefont {J.}~\bibnamefont
  {Bloch}},\ and\ \bibinfo {author} {\bibfnamefont {E.}~\bibnamefont
  {Akkermans}},\ }\bibfield  {title} {\bibinfo {title} {Fractal energy spectrum
  of a polariton gas in a {F}ibonacci quasiperiodic potential},\ }\href
  {https://doi.org/10.1103/PhysRevLett.112.146404} {\bibfield  {journal}
  {\bibinfo  {journal} {Phys. Rev. Lett.}\ }\textbf {\bibinfo {volume} {112}},\
  \bibinfo {pages} {146404} (\bibinfo {year} {2014})}\BibitemShut {NoStop}%
\bibitem [{\citenamefont {Damanik}\ \emph {et~al.}(2016)\citenamefont
  {Damanik}, \citenamefont {Gorodetski},\ and\ \citenamefont
  {Yessen}}]{DamGorYes_inv16}%
  \BibitemOpen
  \bibfield  {author} {\bibinfo {author} {\bibfnamefont {D.}~\bibnamefont
  {Damanik}}, \bibinfo {author} {\bibfnamefont {A.}~\bibnamefont
  {Gorodetski}},\ and\ \bibinfo {author} {\bibfnamefont {W.}~\bibnamefont
  {Yessen}},\ }\bibfield  {title} {\bibinfo {title} {The {F}ibonacci
  {H}amiltonian},\ }\href {https://doi.org/10.1007/s00222-016-0660-x}
  {\bibfield  {journal} {\bibinfo  {journal} {Invent. Math.}\ }\textbf
  {\bibinfo {volume} {206}},\ \bibinfo {pages} {629} (\bibinfo {year}
  {2016})}\BibitemShut {NoStop}%
\bibitem [{\citenamefont {Hofstadter}(1976)}]{Hofstadter_prb76}%
  \BibitemOpen
  \bibfield  {author} {\bibinfo {author} {\bibfnamefont {D.~R.}\ \bibnamefont
  {Hofstadter}},\ }\bibfield  {title} {\bibinfo {title} {Energy levels and wave
  functions of bloch electrons in rational and irrational magnetic fields},\
  }\href {https://doi.org/10.1103/PhysRevB.14.2239} {\bibfield  {journal}
  {\bibinfo  {journal} {Phys. Rev. B}\ }\textbf {\bibinfo {volume} {14}},\
  \bibinfo {pages} {2239} (\bibinfo {year} {1976})}\BibitemShut {NoStop}%
\bibitem [{\citenamefont {Jitomirskaya}(2023)}]{Jitomirskaya_ICM2022}%
  \BibitemOpen
  \bibfield  {author} {\bibinfo {author} {\bibfnamefont {S.}~\bibnamefont
  {Jitomirskaya}},\ }\bibfield  {title} {\bibinfo {title} {One-dimensional
  quasiperiodic operators: global theory, duality, and sharp analysis of small
  denominators},\ }in\ \href@noop {} {\emph {\bibinfo {booktitle} {Proceedings
  of the International Congress of Mathematicians \allowbreak\ 2022}}},\
  Vol.~\bibinfo {volume} {2}\ (\bibinfo  {publisher} {EMS Press},\ \bibinfo
  {year} {2023})\ pp.\ \bibinfo {pages} {1090--1120}\BibitemShut {NoStop}%
\bibitem [{\citenamefont {Thouless}\ \emph {et~al.}(1982)\citenamefont
  {Thouless}, \citenamefont {Kohmoto}, \citenamefont {Nightingale},\ and\
  \citenamefont {den Nijs}}]{ThoKohNigNij_prl82}%
  \BibitemOpen
  \bibfield  {author} {\bibinfo {author} {\bibfnamefont {D.~J.}\ \bibnamefont
  {Thouless}}, \bibinfo {author} {\bibfnamefont {M.}~\bibnamefont {Kohmoto}},
  \bibinfo {author} {\bibfnamefont {M.~P.}\ \bibnamefont {Nightingale}},\ and\
  \bibinfo {author} {\bibfnamefont {M.}~\bibnamefont {den Nijs}},\ }\bibfield
  {title} {\bibinfo {title} {Quantized hall conductance in a two-dimensional
  periodic potential},\ }\href {https://doi.org/10.1103/PhysRevLett.49.405}
  {\bibfield  {journal} {\bibinfo  {journal} {Phys. Rev. Lett.}\ }\textbf
  {\bibinfo {volume} {49}},\ \bibinfo {pages} {405} (\bibinfo {year}
  {1982})}\BibitemShut {NoStop}%
\bibitem [{\citenamefont {Dana}\ \emph {et~al.}(1985)\citenamefont {Dana},
  \citenamefont {Avron},\ and\ \citenamefont {Zak}}]{DanAvrZak_jpc1985}%
  \BibitemOpen
  \bibfield  {author} {\bibinfo {author} {\bibfnamefont {I.}~\bibnamefont
  {Dana}}, \bibinfo {author} {\bibfnamefont {Y.}~\bibnamefont {Avron}},\ and\
  \bibinfo {author} {\bibfnamefont {J.}~\bibnamefont {Zak}},\ }\bibfield
  {title} {\bibinfo {title} {Quantised hall conductance in a perfect crystal},\
  }\href {https://doi.org/10.1088/0022-3719/18/22/004} {\bibfield  {journal}
  {\bibinfo  {journal} {Journal of Physics C: Solid State Physics}\ }\textbf
  {\bibinfo {volume} {18}},\ \bibinfo {pages} {L679} (\bibinfo {year}
  {1985})}\BibitemShut {NoStop}%
\bibitem [{\citenamefont {Osadchy}\ and\ \citenamefont
  {Avron}(2001)}]{OsaAvr_jmp01}%
  \BibitemOpen
  \bibfield  {author} {\bibinfo {author} {\bibfnamefont {D.}~\bibnamefont
  {Osadchy}}\ and\ \bibinfo {author} {\bibfnamefont {J.~E.}\ \bibnamefont
  {Avron}},\ }\bibfield  {title} {\bibinfo {title} {Hofstadter butterfly as
  quantum phase diagram},\ }\href {https://doi.org/10.1063/1.1412464}
  {\bibfield  {journal} {\bibinfo  {journal} {J. Math. Phys.}\ }\textbf
  {\bibinfo {volume} {42}},\ \bibinfo {pages} {5665} (\bibinfo {year}
  {2001})}\BibitemShut {NoStop}%
\bibitem [{\citenamefont {Avron}\ \emph {et~al.}(2003)\citenamefont {Avron},
  \citenamefont {Osadchy},\ and\ \citenamefont {Seiler}}]{AvrOsaSei_phystod03}%
  \BibitemOpen
  \bibfield  {author} {\bibinfo {author} {\bibfnamefont {J.~E.}\ \bibnamefont
  {Avron}}, \bibinfo {author} {\bibfnamefont {D.}~\bibnamefont {Osadchy}},\
  and\ \bibinfo {author} {\bibfnamefont {R.}~\bibnamefont {Seiler}},\
  }\bibfield  {title} {\bibinfo {title} {A topological look at the quantum hall
  effect},\ }\href {https://doi.org/10.1063/1.1611351} {\bibfield  {journal}
  {\bibinfo  {journal} {Physics Today}\ }\textbf {\bibinfo {volume} {56}},\
  \bibinfo {pages} {38} (\bibinfo {year} {2003})}\BibitemShut {NoStop}%
\bibitem [{Note1()}]{Note1}%
  \BibitemOpen
  \bibinfo {note} {One could replace $V_{\alpha }$ by a smooth approximation,
  as in \protect \citep {KelPro_ahp19}. This is useful for irrational
  frequencies, but does not fully resolve the problem for the periodic
  operators.}\BibitemShut {Stop}%
\bibitem [{\citenamefont {Avron}\ \emph {et~al.}(2014)\citenamefont {Avron},
  \citenamefont {Kenneth},\ and\ \citenamefont {Yehoshua}}]{AvrKenYeh_jpa14}%
  \BibitemOpen
  \bibfield  {author} {\bibinfo {author} {\bibfnamefont {J.~E.}\ \bibnamefont
  {Avron}}, \bibinfo {author} {\bibfnamefont {O.}~\bibnamefont {Kenneth}},\
  and\ \bibinfo {author} {\bibfnamefont {G.}~\bibnamefont {Yehoshua}},\
  }\bibfield  {title} {\bibinfo {title} {A study of the ambiguity in the
  solutions to the {D}iophantine equation for {C}hern numbers},\ }\href
  {https://doi.org/10.1088/1751-8113/47/18/185202} {\bibfield  {journal}
  {\bibinfo  {journal} {J. Phys. A}\ }\textbf {\bibinfo {volume} {47}},\
  \bibinfo {pages} {185202, 10} (\bibinfo {year} {2014})}\BibitemShut {NoStop}%
\bibitem [{\citenamefont {Band}\ \emph {et~al.}(2023)\citenamefont {Band},
  \citenamefont {Beckus},\ and\ \citenamefont {Loewy}}]{BanBecLow_MFO23}%
  \BibitemOpen
  \bibfield  {author} {\bibinfo {author} {\bibfnamefont {R.}~\bibnamefont
  {Band}}, \bibinfo {author} {\bibfnamefont {S.}~\bibnamefont {Beckus}},\ and\
  \bibinfo {author} {\bibfnamefont {R.}~\bibnamefont {Loewy}},\ }\bibfield
  {title} {\bibinfo {title} {{MFO} {R}eport: {T}he {D}ry {T}en {M}artini
  problem for {S}turmian dynamical systems},\ }\href
  {https://arxiv.org/abs/2309.04351} {\bibfield  {journal} {\bibinfo  {journal}
  {arXiv:2309.04351}\ } (\bibinfo {year} {2023})}\BibitemShut {NoStop}%
\bibitem [{\citenamefont {Band}\ \emph {et~al.}(2024)\citenamefont {Band},
  \citenamefont {Beckus},\ and\ \citenamefont {Loewy}}]{BanBecLoe_arXiv24}%
  \BibitemOpen
  \bibfield  {author} {\bibinfo {author} {\bibfnamefont {R.}~\bibnamefont
  {Band}}, \bibinfo {author} {\bibfnamefont {S.}~\bibnamefont {Beckus}},\ and\
  \bibinfo {author} {\bibfnamefont {R.}~\bibnamefont {Loewy}},\ }\bibfield
  {title} {\bibinfo {title} {{The Dry Ten Martini Problem for Sturmian
  Hamiltonians}},\ }\href {https://arxiv.org/abs/2402.16703} {\bibfield
  {journal} {\bibinfo  {journal} {arXiv:2402.16703}\ } (\bibinfo {year}
  {2024})}\BibitemShut {NoStop}%
\bibitem [{Note2()}]{Note2}%
  \BibitemOpen
  \bibinfo {note} {The description here is for $\lambda >0$. The tree for
  $\lambda <0$ is a vertical reflected image of the tree described above, see
  details in \protect \citep {BanBecLoe_arXiv24}.}\BibitemShut {Stop}%
\bibitem [{\citenamefont {Bellissard}\ \emph {et~al.}(1992)\citenamefont
  {Bellissard}, \citenamefont {Bovier},\ and\ \citenamefont
  {Ghez}}]{Bellissard1992}%
  \BibitemOpen
  \bibfield  {author} {\bibinfo {author} {\bibfnamefont {J.}~\bibnamefont
  {Bellissard}}, \bibinfo {author} {\bibfnamefont {A.}~\bibnamefont {Bovier}},\
  and\ \bibinfo {author} {\bibfnamefont {J.-M.}\ \bibnamefont {Ghez}},\
  }\bibfield  {title} {\bibinfo {title} {Gap labelling theorems for
  one-dimensional discrete {S}chr\"odinger operators},\ }\href
  {https://doi.org/10.1142/S0129055X92000029} {\bibfield  {journal} {\bibinfo
  {journal} {Rev. Math. Phys.}\ }\textbf {\bibinfo {volume} {4}},\ \bibinfo
  {pages} {1} (\bibinfo {year} {1992})}\BibitemShut {NoStop}%
\bibitem [{\citenamefont {Bellissard}\ \emph {et~al.}(1991)\citenamefont
  {Bellissard}, \citenamefont {Iochum},\ and\ \citenamefont
  {Testard}}]{BelIocTes_cmp91}%
  \BibitemOpen
  \bibfield  {author} {\bibinfo {author} {\bibfnamefont {J.}~\bibnamefont
  {Bellissard}}, \bibinfo {author} {\bibfnamefont {B.}~\bibnamefont {Iochum}},\
  and\ \bibinfo {author} {\bibfnamefont {D.}~\bibnamefont {Testard}},\
  }\bibfield  {title} {\bibinfo {title} {Continuity properties of the
  electronic spectrum of {$1$}d quasicrystals},\ }\href
  {http://projecteuclid.org/euclid.cmp/1104248304} {\bibfield  {journal}
  {\bibinfo  {journal} {Comm. Math. Phys.}\ }\textbf {\bibinfo {volume}
  {141}},\ \bibinfo {pages} {353} (\bibinfo {year} {1991})}\BibitemShut
  {NoStop}%
\bibitem [{\citenamefont {Band}\ \emph {et~al.}(2025)\citenamefont {Band},
  \citenamefont {Beckus}, \citenamefont {Biber}, \citenamefont {Raymond},\ and\
  \citenamefont {Thomas}}]{BanBecBibRayTho_proceedings24}%
  \BibitemOpen
  \bibfield  {author} {\bibinfo {author} {\bibfnamefont {R.}~\bibnamefont
  {Band}}, \bibinfo {author} {\bibfnamefont {S.}~\bibnamefont {Beckus}},
  \bibinfo {author} {\bibfnamefont {B.}~\bibnamefont {Biber}}, \bibinfo
  {author} {\bibfnamefont {L.}~\bibnamefont {Raymond}},\ and\ \bibinfo {author}
  {\bibfnamefont {Y.}~\bibnamefont {Thomas}},\ }\bibfield  {title} {\bibinfo
  {title} {{A} review of a work by {R}aymond: {S}turmian {H}amiltonians with a
  large coupling constant - periodic approximations and gap labels},\ }\href
  {https://arxiv.org/abs/2409.10920} {\bibfield  {journal} {\bibinfo  {journal}
  {arXiv:2409.10920}\ } (\bibinfo {year} {2025})},\ \bibinfo {note}
  {{P}roceedings of ``CA18232: Mathematical models for interacting dynamics on
  networks''. In press}\BibitemShut {NoStop}%
\bibitem [{\citenamefont {Hiramoto}\ and\ \citenamefont
  {Kohmoto}(1989)}]{HirKoh_prl89}%
  \BibitemOpen
  \bibfield  {author} {\bibinfo {author} {\bibfnamefont {H.}~\bibnamefont
  {Hiramoto}}\ and\ \bibinfo {author} {\bibfnamefont {M.}~\bibnamefont
  {Kohmoto}},\ }\bibfield  {title} {\bibinfo {title} {New localization in a
  quasiperiodic system},\ }\href {https://doi.org/10.1103/PhysRevLett.62.2714}
  {\bibfield  {journal} {\bibinfo  {journal} {Phys. Rev. Lett.}\ }\textbf
  {\bibinfo {volume} {62}},\ \bibinfo {pages} {2714} (\bibinfo {year}
  {1989})}\BibitemShut {NoStop}%
\bibitem [{\citenamefont {Kraus}\ and\ \citenamefont
  {Zilberberg}(2012)}]{KraZil_prl12}%
  \BibitemOpen
  \bibfield  {author} {\bibinfo {author} {\bibfnamefont {Y.~E.}\ \bibnamefont
  {Kraus}}\ and\ \bibinfo {author} {\bibfnamefont {O.}~\bibnamefont
  {Zilberberg}},\ }\bibfield  {title} {\bibinfo {title} {Topological
  equivalence between the fibonacci quasicrystal and the {H}arper model},\
  }\href {https://doi.org/10.1103/PhysRevLett.109.116404} {\bibfield  {journal}
  {\bibinfo  {journal} {Phys. Rev. Lett.}\ }\textbf {\bibinfo {volume} {109}},\
  \bibinfo {pages} {116404} (\bibinfo {year} {2012})}\BibitemShut {NoStop}%
\bibitem [{\citenamefont {Kellendonk}\ and\ \citenamefont
  {Prodan}(2019)}]{KelPro_ahp19}%
  \BibitemOpen
  \bibfield  {author} {\bibinfo {author} {\bibfnamefont {J.}~\bibnamefont
  {Kellendonk}}\ and\ \bibinfo {author} {\bibfnamefont {E.}~\bibnamefont
  {Prodan}},\ }\bibfield  {title} {\bibinfo {title} {Bulk-boundary
  correspondence for {S}turmian {K}ohmoto-like models},\ }\href
  {https://doi.org/10.1007/s00023-019-00792-5} {\bibfield  {journal} {\bibinfo
  {journal} {Ann. Henri Poincar\'{e}}\ }\textbf {\bibinfo {volume} {20}},\
  \bibinfo {pages} {2039} (\bibinfo {year} {2019})}\BibitemShut {NoStop}%
\bibitem [{Note3()}]{Note3}%
  \BibitemOpen
  \bibinfo {note} {This interesting observation yet awaits a rigorous
  explanation.}\BibitemShut {Stop}%
\end{thebibliography}%

\end{document}